# Self-doped apical O$^-$ can lead to plane expansion and/or proposed 3-D superconductivity in YBa2Cu3Oy

H Oesterreicher, Department of Chemistry, UCSD, La Jolla, CA 92093-0506

The distance of apical O to planes, d, is introduced as serving as an arbiter in the dispute between different local O coordination (n) in determining charge equilibration in YBa2Cu3Oy. We assign reported cell volume and plane expansion (V+P+) on shot quenching (SQ) from above 600K to O$^-$ connected with (3) at a critical cluster size for increased d, maximizing at 673K and y=6.44. This and a Tc=100K level is ascribed to self-doped apical O$^-$, acting as a plane reducer and expander (P+). By contrast, slow cooling to 260K produces conventional cell volume contracted varieties (V-P-), based on plane oxidizing (4) with its small d=2.3A at a Tc=50K level. V+P+ to V-P- transformation is slow due to electronic rearrangement involving O$^-$, compared to the primarily structural one within V-. A SQ minority at Tc=200K* level and related effects observed on laser pulsing (for y=6.5 Tc=552K*), termed as elevated temperature superconductivity (ETS), are also explained as due to plane expanding n-doping. We assume that central (2) connect p-doped pairs in the bonds of both apical O$^-$ with plane Cu, in line with retrograde increases in Tc with increasing %(2). 3-D superconductivity is a result of d now being comparable to plane dimensions.

## 1. Introduction:

### 1.1. New challenge in understanding YBa2Cu3Oy=6.5 in the wake of elevated temperature superconductivity

The discovery of signs for elevated temperature superconductivity (ETS), induced as a fleeting state through short infrared laser pulses [1,2], has posed new challenge to understanding cuprates. A conventionally prepared YBa2Cu3Oy=6.5 ceramic with Tc=52K was driven to one showing the hallmarks of coherent interlayer transport reminiscent of superconductivity up to Tc=552K* (star denotes ETS). The electronic transition is accompanied by a structural one in which the 5-fold O coordinated plane copper, denoted (5), become more separated within their double layers by two picometers. As the plane expands, d the distance of its Cu to the apical O decreases. Also, the double layers become more buckled and the layer distance between them becomes thinner. It has been suggested that this unusual state is arising from pairs also on chains [3]. A question naturally arises whether similar principles can also be obtained in more stable states.

In order to answer this question we survey phenomenology obtained on shot quenching. This technique opened new insights into the possible charge balances near the nominally un-doped semiconductor. It added over-looked structural electronic property relations during the initial explorations such as strong cell volume and plane expansions (V+P+). We include signs for minority Tc=200K* (extrapolated for un-substituted from partially Ni substituted material with Tc=150K) in specially quenched materials [4] as further examples for ETS, and explain them as originating from plane n-doping and (2) being the center of the corresponding p-doped pairs on the bonds of apical O with plane Cu. Accordingly one expects a retrograde dependence with O content according to Tc=%(2)const, as observed for the photo-induced reflectivity edge.

For understanding these phenomena in a material context, the role of competing d of local chain environments is investigated. We will in the following refer to different local O coordination (n) on chain sites and the conventionally prepared orthorhombic based on chain sites (4) and (2) as O42 structure (orthoI) and a high temperature modification based on (3) as O3. O243 corresponds to transitional ladder structures between those boundary phases with non-doping (3) as a minority in the O24 derived V- material of small d. By contrast, O324 maintains elements of O3 such as V+P+, increased axial ratios or d, being based on the presence of n-doping (3) as combined with O$^-$. This sets the stage for the relevant competitions in d(n).

O3 has been obtained through shot quenching or slow re-oxygenation of specially treated materials (973K anneal) at 673K. It is identified through TEM, NQR or absence of O2 evolution in acid. It can have cell volumes and plane metric far exceeding the one of the respective semiconductor (V0) and be located roughly symmetrically of it from V- [5]. Its survival at room temperature without noticeable transformation to V- indicates a different electronic state, the transformation of which is requiring considerably higher activation energies.

ETS phenomenology will be compared with the observation of bulk superconductivity at a 100K level in these cell volume expanded (V+) materials of O3 structure [5-11]. Also of interest is a non-superconducting c-axis contracted phase, denoted C- with a=c/3=3.87A [12, 13], displaying somewhat similar expanded plane characteristics as a member of a thick family of compounds concerning its axial ratios. Cell volume is characteristic of the semiconductor (V0). P+ effects are hinting at plane n-doping as well, which is hindered from becoming superconducting by doping competition on chains. However, we will point out that this unusual structure could potentially be modified to fulfill requirements for ETS and become of pivotal importance for understanding relevant phenomena. For a general understanding of the energetic of relevant structure types one has to consider O-O interaction potentials [14-26]. These studies have so far not included V+P+ types.

Some of these materials can rapidly transform to the conventional ones in several steps. The abrupt onset of some transitions with composition hints at the presence of criticalities (e.g. 600K as temperature where V+P+ principles originate). An accounting of the phenomenology is here attempted in terms of criticalities in d of the local O coordination (n) of the chain site with n=2, 3 and 4, as d is depending on the relative concentration of (n). In some transition regions Tc can change within small changes of parameters such as cell volume. This reflects an additional electronic origin of the transition rather than a purely structural one. This origin will be identified as arising from $O^-$. Accordingly each of the three (n) on the chains has a range where it can assert its own type of superconductivity as a special doping type. This lends generality to the remarkable phenomena.

In order to understand driving forces for different doping types we outline conditions for n-doped planes as depending on high temperature synthesis where the rigid planes can satisfy their increased space demands. As planes, according to tolerance factors, will attract space-demanding electrons, the chain site will have to reciprocate and be oxidized and relatively shrinking. Chain Cu will be drawing closer the negative apical O (smaller d1) and so removing it from the planes and increasing d. The size of d can therefore in competitive situations gate the doping type and act as its fingerprint. Electron doping on planes is usually reserved for systems without apical O that place the out of plane O in locations between plane Cu and so tend to enlarge planes. While systems with apical O are not tending to be electron doped on planes at room temperature, this exclusionary principle appears to be relaxed on quenching from high-energy environments, where the rigid planes have higher space demands. In a further conceptual step for laser pulsing we consider that the self-doping on planes, of expanding electron type, can draw the apical O close enough to the planes for direct bonding. This new situation allows the establishment of a balancing bond order of holes on chain (2) that can be transformed into pairs and introduce 3-D coupling within the plane-chain-plane (PCP) sandwich. In this sense the deleterious effect of the apical O will have been turned to advantage. This usually weak bond, or the corresponding hybridization with the plane, is now transformed into a hole doped bond on either side of the chain sticks touching the planes and constituting additional pairs. It is also possible that apical O of $O^-$ nature is directly involved over metal-ligand charge transfer in the mechanism of ETS.

It is known that shortening d, can take a deleterious influence on the magnitude of Tc through hybridization with p-doped plane Cu [27, 28] where pairing has been assumed to take place. In fact, algorithms can use d for predicting Tc [29-32] for n and p-doped planes. These relations obviously will not hold for ETS with its presumed especially low d at high Tc. This low d, established by (2), has to draw in (4) as well, and keep it in an oxidized doping inactive state. We will also delineate how the influence of a competition between local environments of chain Cu in its various O coordination (n) can lead to criticalities in this respect [32]. As an example, at moderately high concentration, (4) can act as a plane oxidizer. Conversely, (3) at high concentration can act as plane reducer [5]. It is possible that (3) may also act as a plane oxidizer when forced by the local constraints of majority (4). The phenomenology of the spatial influence that different (n) can take on each other is discussed in terms of a possible collapse of the distances of a minority local component into a majority components at low minority compositions. The result can be either localization catastrophes or the mutual support of doping. In the following we will introduce developments leading to the present understanding.

### 1.2. Tc predictions and distance of the apical O to the cuprate planes

Elevated temperature superconductivity in cuprates is here defined as corresponding to Tc beyond a theoretical limit of about 150K. This limit is indicated for predominantly 2-D superconductivity in the plane isolation model [29-31] with

equation1 $Tc = ic$,

where the optimal plane isolation factor $i=600K$, and $c=0.25$ appears to be the optimal doped charge concentration in plane p-doped cuprates. The model is general and extendible to other classes such as pnictides [31]. We assume that doped charge concentration actually represents paired charge concentration over a large range before over-doping for systems such as YBa2Cu3Oy. Here the dopant chain site is relatively ordered and self-doping. This includes part of the region of insipient orthorhombic splitting and manifests itself in a calibrated relationship with changes in lattice parameter c or parameter d. For YBa2Cu3Oy=6+z, the relatively close approach of the apical O to the plane with the important distance d, diminishes the isolation factor to about $i=400K$ on bond valence calculations or an empirical $i=73d^2$ with d in A. While a dependence of Tc on d was suggested independently, based on hybridization calculations [27, 28], no predictive scheme was developed as a result. The linear range of Tc with doping over a large range in doped charge c or y is borne out by experiment although special preparation can also lead to Tc plateaus through the incorporation of varying amounts of non-doping (3) in O42. These Tc plateaus reflect the magic pair concentrations of $c=0.167$ and $c=0.222$ respectively, derived as follows on a bond order model.

It is possible to relate charge concentrations c directly to the concentration of active dopant. For the O42 structure we assume linear doping in a range up to optimal at $c=0.22$. This is reached at a bond order with pair repeat distances 3a0x3b0 and $c=2/3x3=0.22$, while $c=0.167$ corresponds to 2/3x4. $c=0.222$ occurs at %(4)= 89% at y=6.89, reducing equation1 to $Tc = \%(4) \times 1.0 = ic$ where 1.0 is a constant. For the n-doping (3) a similar relation $Tc = \%(3) \times 1.1$ holds with a slightly different constant. This relation also optimizes at $c=0.22$ or y=6.44, as expected from bond order principles. Both relations are based on the assumption of direct proportionality over some range of superfluid density and dopant concentration. Similar formulations should therefore extend to ETS according to $Tc = \%(2) const$. From available data we derive const=11.

The distance of the apical O to the cuprate planes, d, will be taken below as mediator between doping types. Neutrons documented the response of d to y of conventionally prepared materials [33]. Uniform values of d for the chains are reported although some local buckling could occur. It appears that d stays relatively constant with y near both boundary phases. In particular, the semiconductor keeps nearly constant d=2.46A over a wide range in y. With the incipient tetragonal-orthorhombic phase transition, d strongly declines and thereafter shows more mildly declining values that extrapolate roughly between the values of the parent semiconductor and the superconductor at y=7 near d=2.23A. The distance of chain Cu to the apical O, d1 shows a reciprocal if less pronounced behavior with d=1.78A for y=6.0 and increasing over a range in y stabilizing on approach of y=7.0 near d1=1.82A. This tendency in d1 reflects the expected general reciprocity between contracting planes and expanding chains, which is not reflected in d. However, the tighter the bond between chain Cu and apical O, d1, the longer one expects the bond to be between apical O and plane Cu, d. As d contributes stronger to the overall distance d1+d, one will not expect c-axis expansion but rather a contraction on p-doping the planes. An exception, where a straightforward reciprocity is expected to hold, is in cases of pair formation on chains proposed for ETS. The phenomenology of the spatial influence that different (n) can take on each other has been discussed in terms of a competition model [32] with the possible collapse of the distances of a minority local component into a majority components at low minority compositions.

### 1.3.a Hallmarks of the quench phenomenology

It became clear early on that O42, observed on oxidizing slow cooling, was a low temperature modification that could change slightly at moderately higher temperatures to modifications involving (3) at constant y. As a result transitions from lower Tc to higher values were routinely observed on room temperature annealing of moderately fast quenched materials [7-10, 34-38]. Changes in Tc are largest close to the onset of superconductivity, as the relative potential change of (3)/(4) is largest. While these studies were rather extensive they missed some of the most intriguing aspects of phenomenology at >600K that were obtained by fine tuning quenching technique and by engineering increases in activation energy of the transition reaction through chemical means (partial substitutions). Shot quenching with small sample size of 0.2g shortened quenching time to 0.1s. Indications for strong cell volume and plane expansions were a result for several rare earth and Y analogs near experimental y=6.46, which was assumed to reflect an ideal plane bond order at y=6.444 with c=0.222=2/3x3. The V+P+ effects started to pick up at quenching temperature Tq=600K, corresponding to the subperoxide stability belt [39], and maximized at 670K with volume increases of up to 2.7% over V-. In some instances they were coupled with a doubling and quadrupling of Tc. Minority components with Tc=200K* level are here related to ETS.

These studies were stimulated by ones of the effects of partial transition metal substitutions for Cu. The resulting micro-cluster formation [40-46], leading to some reductions in Tc and rise in magnetic remanence or intrinsic hardness, was also found to reduce the speed of transformation reactions from the metastable high temperature O3 modifications to the conventional O42 orthorhombics. By utilizing this slowing of transformations, synthesis and monitoring of fast transitions was facilitated. The main part of this work involved $YBa_2Cu_{3-x}Ni_xO_y$ with x<0.05. While strong cell volume expansions were found also with x=0 and with several rare earth analogs, the increases in Tc when calibrated to $YBa_2Cu_{3-x}Ni_xO_y$ with x=0.0 were so far only found with the Y analog. In $YBa_2Cu_{3-x}Ni_xO_y$ considerable complexity is encountered concerning the response to oxygenating (low temperature equilibration) or reducing preparations (shot quenching) termed Ko phenomenology. Besides increases in Tc, also absence of Tc at high y is observed and these materials show large amounts of (3) and resistance to re-oxygenation. In fact, Ko phenomenology can be attributed to the competition of local environments of (3) and (4) with their, under certain circumstances, inherently different doping mechanisms based on $O^-$ or straightforward ionic metal charge storage respectively. Thermodynamic studies have dealt with the question of stability and O-uptake [47-51]. Presence of $O^-$ was documented by spectroscopy [52].

In summary it appears that the mercurial O half filling situation near $YBa_2Cu_3O_{6.5}$ can display a variety of Tc levels. Conventionally prepared, materials display a Tc=50K level but shot quenching raises this to 100K (with indications for 200K*, or 0K in the transitional range) and results from laser pulsing show now a Tc=550K* level. We will therefore draw on the changeable nature of the local environment at O half filling and the variety of self doping levels in bond order structures, both on planes and chains, for explaining the corresponding Tc level structure. The general nature of this bond or charge ordering in electronic crystals has been seen in STM and ARPES [53,54] where indications for pairs at 3a0/2 location were observed in several charge-lattice commensurate bond order patterns on planes. Such pairs at 3a0/2 should in principle also be a possibility on chains and can be expected to order themselves into the plane-plaid bond order.

## 2. Results

We will now use available literature data to construct an understanding of the way temperature can act as a variable self-doping agent through the gating influence of d in YBa2Cu3Oy near O half filling (123/6.5). Accordingly local environment is responsible for 3 types of doping and these can be obtained through various modes of fast quenching or slow cooling to room temperature. To begin with, the transition of O42 to O3 types with rising temperature is considered to take place initially in infinitely adaptable superstructures, O42 being obtainable in its pure form below 260K and O3 being documented after shot quenching from 673K (bond order principles dictate an ideal y=6.44 so that an actual designation would be O32 for the material in question). We hold that disorder will play a minor role, given the energetic cost of strong O-O repulsion potentials. Accordingly chain length of (4) decreases systematically from infinity to 0 by simple shifts in the O position. As an example chain length is shortened by about half on moving one O out of a chain of (4) into the chain of (2) as the temperature is raised beyond the characteristic temperature for the given chain length. In the process the chain of (4) is now locally transformed into environment (4)O(3)(3)O(4) and the transferred O creates a new environment within the corresponding chain of (2) of (2)(3)O(3)(2). For understanding the active local redox processes it suffices to denote the first local environment as (4)O(3) and the second as (3)O(3). Further below we will consider (4)O(3) as a plane oxidizer and (3)O(3) as neutral or as a plane reducer, dependent on arguments concerning magnitude of d as resulting from different clustering of (3)O(3). We denote the self-doped oxidized plane as $(5)^{ox}$ and consider it charge balanced by $(4)^{Red}$.

Conversely on cooling from 673K one will encounter the change from O3 to O243 by a similar if much slower process, as it involves an electronic rearrangement in addition to a structural one. An ideal well ordered O3 has infinite motives of (3)O(3)(3)O(3). When shot quenched it is considered to develop self-doped $O^-$, perhaps involving subperoxide subunits through processes of tension self-doping in which misfits between plane and chains are mediated. This can act in increasing cell volume and plane metric far beyond the corresponding semiconductor, as well as chemically reducing planes and so establishing n-doped superconductivity of the Tc=100K level. On cooling, this motive will be transformed along two parallel orthorhombic rows of (3)O(3)(3)O(3)(3)O(3) and (3)(3)O(3)(3)O(3)(3) by moving the central O from the first row to the second. This leads to (3)O(3)(2)(2)(3)O(3) and (3)(3)O(4)O(4)O(3)(3) respectively. Locally this means the destruction of attendant apical $O^-$ and a switch in doping. The presence of (4), as a plane oxidizer, will now be in competition with regions of the potential plane reducer (3)O(3), leading eventually to the loss of n-doped superconductivity through doping competition. For understanding these processes we will now consider attendant changes in local environment as reflected in d.

For orienting purposes, distance levels of d for different local O coordination (n) on chain sites in YBa2Cu3Oy are given in Table1. The semiconductor near y=6.0 supplies data for non-doping d(3) and d(2) in semiconducting 2+ and 1+ ionic state both of order 2.5A. d(4)=2.3A for plane p-doped superconductor near y=6.5 with oxidative preparation through slow cooling to 260K. On increasing the shot quenching temperature, starting around 600K, the more enigmatic cell volume expanding states involving (3) are increasingly encountered. We ascribe them to the presence of a critical cluster size such as a minimum of 2 adjacent (3)O(3) configurations, inducing the generation of $O^-$. At Tq=673K and y=6.44, a V+P+ variety is built solely on (3) and (2), where (3) is considered as n-doping the planes. For this material we estimate d(3)=2.7A as symmetrically disposed from d(4)=2.3A. For quenching from <600K we assume d(3) to compete with d(4) at around 2.3A.

A summary of structural and electronic types in YBa2Cu3Oy near O half filling, obtained by shot quenching and slow cooling is shown in Table2. Materials are ordered with respect to quench temperature, Tq. This parameter can be seen as a variable self-doping agent. With preparation through slow cooling to 260K the orthorhombic alternate chain structure O24 (or OI) forms, as based on (2) and (4). With rising temperature non-doping (3) is generated according to equilibrium1 (4)+(2) =2(3) being shifted increasingly to the right. This has been assumed to take place in ordered infinitely adaptable ladder modifications that exhibit only slightly increasing cell volumes up to 600K. Beyond an extrapolated onset of 600K cell volumes strongly increase. In this range we assume the generation of self-doped $O^-$ in connection with (3) due to a shift in the equilibrium between metal and ligand oxidation where $O^-$ is referred to as $L^-$ in the metal-ligand notation. From 600K to 670K therefore mixed V- and V+ states occur, where the V- part corresponding to O243 quickly transforms to O24 regions. The V+ part, however, is rather stable at lower temperatures to transformation from O32 to O324. The part having transformed on to O243 then rapidly transforms to O24.

The above extends conventional concepts to new situations. However, in considering ETS, altogether new concepts are needed. We proceed on the assumption that additional pair formation on (2) can explain aspects of the unusual effects. Such additional pair formation is in line with a bond order of pairs in 3a/2 position on chains, corresponding to doped bonds between Cu and O in trijugate position, as observed on the planes. Remarkably, the chain system is similarly capable of supporting such doped trijugate bonds. Such a situation can be expected to support 3-D superconductivity. Accordingly we assume collapse of d(2) into d1. This then corresponds to plane metrics so that d(2)=1.95 to 2.0A. This collapse will dramatically shorten the c-axis as observed on laser pulsing (the corresponding observed minority Tc=200K* phase has not been identified yet). (4) is considered to be forced to stay localized and doping inactive in this high energy environment. The related C- phase represents yet one more variation on this c-axis contraction theme.

We now delineate selected kinetic data from the literature for transformations amongst various types in order to support the idea that plane n-doping and V+P+ effects are connected with doped $O^-$. In essence we quantify the argument that the transition observed near room temperature within V- has to be much faster than the one encountered in the synthesis at elevated temperatures. As the former is known to be primarily of structural nature we consider the latter to be different in that transformation is slowed primarily by electronic rearrangement. We propose that this involves $O^-$. This is expected to be germane to related high temperature phases.

Changes in structural parameters or Tc have been determined, following $t=Ae^{E/kT}$ [7]. The transition within V- corresponds to materials initially prepared by slow cooling. Samples are then shot quenched in the temperature interval 320-473K. Phenomena saturate for >473K as at this temperature the shot quench has t=0.1s, limiting observation of this type of transformation at higher temperatures because there they would be too fast to freeze in. Values of E=92kJ are typical, as is A=1.0x10$^{-12}$s. For 298K t=300min is representative. One also notes a similarity of E with values of O diffusion, indicating that these processes are diffusion controlled and mainly structural in nature.

Similar phenomenology is also observed at face value for materials shot quenched from a limited range >600K and measured at 298K. Values of E and A are comparable to the above. Partial Ni substitution according to YBa2Cu3-xNixOy with x=0.020 can increase E to 96kJ, slowing this reaction. One part of the transformation therefore corresponds to the structural shift in equilibrium1 from (3) to (4) and (2). However, the transformation keeps the telltale signs for its different parentage relatively intact, such as increased V or axial ratios c/p=1.015 with p=3(a+b)/2, as a family property. This hints at an incomplete transformation to ideal V-. A second very slow transformation stays accordingly hidden. We assume that materials with Tq>600K are partly based on $O^-$ units, which have slower transformation kinetic to their metallic ionic counterparts. In fact, their survival to room temperature indicates that their t>0.1s for Tq>600K, otherwise they would have been erased during SQ. We denote this earlier and intrinsically much slower reaction as transition V+p/V+. In this notation V+p stands for a parent compound of V+ that need not have its properties but develop them during SQ as a slow electronic rearrangement. V+ is optimized around 673K, while it is in a mixed state from 600-673K. It can survive to 950K.

While we do not know the nature of V+p, we assume it to be stable around 673K in O32 form for ideal y=6.44. It probably already contains doped $O^-$ in the form of subperoxide units, which on quenching can rearrange their structural and electronic features to plane n-doping of the self-doping type. These initial transformation processes V+p/V+ are therefore charge transfer controlled and intrinsically much slower. We attempt now to obtain an estimate for E for the transition O32/O24. When a conventionally prepared material is quenched into liquid N2 from 1080K and subsequently air annealed at 950K for 10 min, followed by SQ, a relative V expanded orthorhombic is observed, while after 2 h at 950 K a cell volume contracted orthorhombic obtains. From the latter we assume t=1h at 950K for the transformation from V+ to V- at 950K. Assuming A as constant, one calculates accordingly E=280kJ from $\ln t = \ln A + E/kT$. This type of electronic rearrangement therefore will be dramatically slower at room temperature, producing values of years rather than min for t. In one case V+ was observed to have survived unchanged for a year.

The situation is more complex for Ta>950K. Examples are: The Tq=1080K material held for 10 min at 970K does not produce resolved X-ray diffraction peaks, testifying to a transitional crystallographic state. Initial calcining at 1053K leads to the C- phase of T3 type. The preparation at 970K leads to lack of x-ray diffraction patterns, possibly due to a transition between relatively unrelated C- phase of T3 parentage and O3. It is also possible that amorphization is a result of the inherent metastability vis-à-vis decomposition into metal oxides and peroxides of the form $BaCuO_{2.7}$. In situ studies indicate the boundary between T and O at 940K although this appears uncertain due to the possibility of slow transformation reactions.

## 3. Discussion

The aim of this study is to analyze existing literature data for their implication concerning the potential for unusual doping types of which the recent laser pulsing with signs for Tc=552K* in $YBa_2Cu_3O_y$ near O half filling is an example. Materials that are structurally related to the laser pulsed ones with respect to P+ effects can also be synthesized by shot quenching. They also show signs for elevated Tc and unusual doping types. Their genesis and properties are therefore of special interest for comparisons.

Generally, 123/6.5 materials appear unremarkable on increasing temperatures when studied in situ, indicating only a gradual shift to structural properties of the semiconductor. However, near an ideal bond order at y=6.44 (c=0.22=2/3x3), materials can gain unusual doping properties on shot quenching. We assume that this is happening through the buildup of internal misfit between plane and chains on quenching. It can be seen as a counterpart to the temporal excitement through laser irradiation. As both processes involve high energy exposure they are leading to expansion of the unaccommodating rigid planes presumably through n-doping. This corresponds to a basic new concept.

For further understanding the effects of laser pulsing it can be illuminating to investigate the shot quenched materials with respect to structural transitions in the process of reordering kinetic. It appears that an intrinsically slow process lies at the heart of the O3/O24 transformation during shot quenching. For this we hold responsible a slow electronic rearrangement from $O^-$ to metal ions as charge carriers. By contrast, the conventional type of kinetic of aging at room temperature involves predominantly a structural one from O243 to O24. The latter essentially corresponds to the gradual transition from states with large portions corresponding to semiconductor components (3)O(3) of T3 to the well-ordered superconductor based on (4) and (2).

However, here we propose that larger aggregates of (3)O(3) in O3 are not producing a semiconductor. Rather, micro-clusters involving (3) can lead to strong V and plane expansion far beyond the semiconductor. We ascribe this to unusual n-doping on planes and suggest that it depends critically on the size of micro-clusters involving (3) with (3)O(3)(3)O(3) as a minimal motive for generating $O^-$. From this we can also infer that similar processes involving n-doping can lead to ETS. However, while for V+ one doped apical $O^-$ is assumed to be associated with (3) along the c-axis, for ETS we postulate that both apical O around (2) house holes as pairs in their bonds to plane Cu. The result is c-axis contraction and 3-D superconductivity, denotable as P+C-.

Generally, the windows where unusual effects are observed remind of classical peroxide stability belts. However, as in situ no V+ effects have been observed, this indicates that stress is needed for the generation of $O^-$, which then is difficult to transform into modifications based primarily on metals as a charge balance. Subperoxide doping through stress expands the notion of peroxide stability belts. In this respect one notices that 123 materials are thermodynamically stable in the semiconducting state at synthesis conditions but become metastable on cooling, with the chance of further O uptake, vis-à-vis peroxides such as Ba Cu O2.7 and metal oxides [50]. This has to do with their increased overall oxygen uptake over the superconductor.

Another exception to decomposition, at higher temperature, is the C- phase obtained after initial calcining at 1053K. It may also contain subperoxidic $O^-$ at relatively high y, showing the hallmarks of n-doping in strongly expanded plane metric. In this sense one can consider V+ and its derivatives within O324 to represent one in a series of $O^-$ containing materials. Generally, the rarer n-doping in cuprates requires ionic doping with a higher valence metal ion. Here we propose an alternative where a potential dopant such as (3)$O^-$ adduct can become self-doped by becoming partly oxidized. For our case, it is not surprising then that the $O^{2-}$ ion can fulfill this self-doping role in the peroxide temperature belt of stability. To play this role, the $O^{2-}$ ion has to be associated with a mercurial metal ionic state such as with (3) together with the usual large alkaline earths like Ba. n-doping accordingly is readily available for 123 under conditions of stability of $O^-$. This can serve as a hint for the origin of ETS as well.

It is of interest that the conventional n-doped 214 materials [55] also have large plane metric and become superconducting on quenching from elevated temperatures. They do however not contain apical O and are not self-doped. This makes the presence of $O^-$ as the reducing dopant on the charge balance layer less likely. Other 123 materials that may show signs for ambipolar doping between p and n-doping include (YLa)(BaLa)2Cu3Oy [56], which would be similarly based on doping by a higher valence metal ion (BaLa) and not on self-doping.

The state of $O^-$ is presently not clear although one can assume some O self-bonding in extended units, minimizing local distortions. We will accordingly speculate on the actual nature of subperoxide units $O_q^{1-2q}$. For V+ at y=6.44 we assume pair bond order on planes with repeat pattern of 3a0x3b0. Accordingly doped charge concentration for one pair is c=2/9=0.22. Within this kernel a corresponding charge pattern has to develop on chains. Instead of now letting (3) carry the doped charge as ionic $Cu^{3+}$ in a sea of $Cu^{2+}$, we search for arrangements within the chain related O. As there exist only ½ of b-axis O per (3), we assume the apical c-axis O is involved in O3. This is also born out by its influence on expanding the c-axis. If we consider that the subperoxide formation extends to the nearest 2 apical O forming a charge resonating subunit of 3 with the b-axis O, then this unit is depicted as $O_3^{-5}$. Other arrangements are possible. The electronic nature of these subperoxides is not known but it is possible that it has aspects in common with solvated electrons or electrides [57,58] where charge is dislocated over a structural cage.

A problem remains concerning the analytic prediction of Tc in the plane isolation model. We note that subperoxidic $O^-$ is characterized by high d. In the respective formalism d=2.7A comes close to Hg based analogs. Accordingly these systems should be based on largely isolated planes with high Tc. So far we have taken the signs for a doubling of Tc around y=6.44 as due mainly to a doubling in pair concentration. As the changed isolation factor accordingly would not come into play one might hold the need for a downward Tc recalibration for n-doping as responsible. It is however possible that the increase in Tc directly reflects the increased plane isolation. In this case a new interpretation for the respective pair concentration would be in order.

We now attempt a conceptual expansion in which we expect (2) to potentially act also as a plane reducer. This could be triggered by the stress induced by shot quenching from a high energy environment where the rigid planes tend to expand. However, in this case the already low d(4) is considered to collapse together with d(2) into comparable metrics to the ones on planes. This has been suggested to be the case for ETS where an additional pair formation around (2) of p-type has been assumed as leading to high superconducting carrier concentration and 3D effects. This adds a 3$^{rd}$ and altogether new type to doping, besides conventional plane p and n-doping phenomenology. Based on the plane expansions, ETS can be seen as related to plane n-doping. Laser pulsing has indicated that states are available upon high energy exposure that involve plane expansion and c-axis contraction in YBa2Cu3Oy near O half filling.

For V+ the high temperature generator is O3, and this may also pertain to the minority component of ETS type in shot quenched material with Tc=200K*. However, we propose that T3 could also produce similar varieties in the presence of some (2). In this way each of the 3 structure types of Stripes, Argyle or Herringbone would have its special doping just as each of the 3 local chain configurations has. Underlying these rules is the mercurial distance to the apical O, depending on local chain environment. This can lead to changes in doping types, potentially involving criticalities concerning the location of the apical O from the planes, d.

**4. Conclusions**

Amongst the 3 different local chain O coordination (n) in YBa2Cu3Oy near O half filling each is proposed to have its own doping type and resulting superconductivity. Conventional slow cooling continuously transforms successive high temperature phases of T3 and O3 type to plane p-doped O42 with metal (4) as self-doping oxidizer (d=2.3A) resulting in P-V- effects. By contrast, shot quenching in the range from 670-950K can lead to P+V+ effects, indicating a doping flip to plane n-doping. Accordingly at sufficient concentration of O3 component, (3) can act together with its ligand O$^-$ as plane reducer in the peroxide stability range (estimated d=2.7A). This results in plane expansion and, in some cases, a doubling or further multiplication of Tc. Another distinguishable high temperature state is the C- phase with P+C- phenomenology, also suggesting the involvement of subperoxide units $O_q^{1-2q}$, with possible relevance to ETS. Survival of these states to 298K indicates slow intrinsic transformation kinetic from P+ to P-, even at elevated temperatures, due to the need for electronic rearrangements. In a similar vein it is proposed that elevated temperature superconductivity can arise from additional hole-pairs on chains, with (2) and its 2 apical O$^-$ as a likely pair carrying candidate. In this case, d is assumed as comparable to plane metric allowing 3-D superconductivity.  This results in retrograde Tc increasing with %(2). Besides on laser pulsing, ETS appears to occur in special shot quenched preparations in a Tc=200K* level minority phase. The concept of subperoxidic O$^-$ is considered as a common origin of unusual n-doped varieties such as V+P+ and ETS.

**References**

1 S Kaiser et al, Phys. Rev. B 89, 184516 (2014)
2 R Mankowsky et al, Nature 516 71 (2014)
3 H Oesterreicher, arXiv: 1503.00779 (2015)
4 D Ko et al, Materials Research Bulletin 29 1025 (1994)
5 D Ko et al, Physica C: Superconductivity 277 95 (1997)
6 D Ko et al, Materials Research Bulletin 231 252 (1994)
7 H Oesterreicher et al, Journal of Alloys and Compounds 306 96  (2000)
8 JR O'Brien et al, Journal of alloys and compounds 267 70 (1998)

9 JR O'Brien et al, Journal of solid state chemistry 135 307 (1998)
10 HS Kim et al, Journal of Alloys and Compounds 339 65 (2002)
11 JR O'Brien et al, Physica C: Superconductivity 388 379 (2003)
12 A Manthiram et al, Nature 329 701 (1987)
13 H Oesterreicher et al, Materials Research Bulletin 23 1327 (1988)
14 G. Ceder et al, Phys. Rev. B., 44 2377 (1991)
15 HF Poulsen et al, Phys. Rev. Lett., 66 465 (1991)
16 R McCormack et al, Phys. Review B., 45 12976 (1992)
17 DJ Liu et al, Phys. Rev. B., 52 9784 (1995)
18 P Schleger et al, Phys. Rev. Lett. 74 1446 (1995)
19 E Straube et al, Physica C., 295 1 (1998),
20 G Castelani et al, International J. of Modern Physics B., 13 1073 (1999)
21 NH Andersen et al, 37 Physica C., 317 259 (1999)
22 F Yakhou et al, Physica C., 333 146 (2000)
23 R Liang et al, Physica C., 336 57 (2000)
24 R Liang et al, Physica C., 383 1 (2002)
25 M Zimmermann et al, Phys. Rev. B. 68 104515 (2003)
26 R Liang et al, Phys. Review B., 73 180505 (2006)
27 Y Ohta et al, Phys. Rev. B 43 2968 (1991)
28 C Di Castro et al, Phys. Rev. Lett. 66 3209 (1991)
29 H Oesterreicher, Solid State Communications 142 583 (2007)
30 H Oesterreicher, Physica C: Superconductivity 460 362 (2007)
31 H Oesterreicher, arXiv:0811.2792 (2008)
32 H Oesterreicher, Materials Research Bulletin 30 987 (1995)
33 RJ Cava et al, Physica C. 156 523 (1988)
34 C Claus et al, Physica C 171 205 (1990)
35 JD Jorgensen et al, Physica C 167 571 (1990)
36 BW Veal et al, Phys. Review B 42 6305 (1990)
37 J Kircher et al, Phys. Review B 48 9684 (1993)
38 H Shaked et al, Phys. Review B 51 547 (1995)
39 H Oesterreicher, Journal Solid State Chemistry 136 390 (2002)
40 MG Smith et al, Physical Review B 42 4202 (1990)
41 MG Smith et al, Journal of Applied Physics 69 4894 (1991)
42 LT Romano et al, Physical Review B 45 8042 (1992)
43 MG Smith et al, Physica C: Superconductivity 204 130 (1992)
44 MG Smith et al, Journal of Solid State Chemistry 99 140 (1992)
45 H Oesterreicher et al, Journal of Alloys and Compounds, 269 246 (1998)
46 H Oesterreicher, Applied physics 15 341 (1978)
47 H Oesterreicher et al, Materials Research Bulletin 22 1709 (1987)
48 H Verweij, Ann. Phys. Fr., 13 349 (1988)
49 H Oesterreicher et al, Materials Letters 6 254 (1988)
50 H Oesterreicher, Journal of alloys and compounds 267 66 (1998)
51 H Oesterreicher, Journal of Low Temperature Physics 117 993 (1999)
52 S Horn et al, Phys. Rev. B 36, 3895 (1987)
53 T Hanaguri et al, Nature 430 1001 (2004)
54 KM Shen et al, Science 307 901 (2005)
55 NP Armitage et al, arXiv:0906.2931v2 (2010)
56 K Segawa et al, Phys. Rev. B 74(10), 100508 (2006)
57 H Oesterreicher et al, Journal of Solid State Chemistry 1 10 (1969)
58 JL Dye, Science 301 607 (2003)

Table 1. Distance levels of d of apical O to plane Cu in Angstrom for different local O coordination (n) on chain sites in YBa2Cu3Oy. e stands for estimated. The semiconductor near y=6.0 supplies data for non-doping d(3) and d(2) in 2+ and 1+ ionic state of 2.5A. d(4)=2.3A is for plane p-doped superconductor near y=6.5 with preparation through slow cooling to below 300K. On shot quenching these materials equilibrium1 (4)+(2) =2(3) is shifted to the right and Tc decreases, showing increasing amounts of non-doping (3) with structurally unremarkable changes. Upon SQ from >600K strong cell volume increases set in. They are associated with generation of $O^-$, optimizing at 670K. At 670K and ideal y=6.44, a V+ variety is based on (3), n-doping the planes and leading to a Tc=100K level. From the V+ effects we estimate d(3)=2.7A. A minority component exhibits a Tc=200K* level. This is taken as elevated temperature superconductivity, which we have explained as a result of reducing plane doping, leaving hole pairs on bonds between apical O associated with (2) and plane Cu, doping electron pairs on (5). As a result d collapses into plane metric with 2.0A. ETS also occurs on laser pulsing of O24 where it is characterized by strong c-axis contraction and plane expansion so that d and the corresponding intra-planar spacing can become comparable in these 3-D superconductors. (4) is considered to be forced to stay localized and doping inactive in this high energy environment.

| d(4) | d(3) | d(2) | doping type |
|------|------|------|-------------|
| 2.0e | 2.5 | 2.5 | non-doping |
| 2.3 | (2.3e) | - | oxidizing doping |
| - | 2.7e | | reducing doping |
| | | 2.0e | ETS doping |

Table 2. Structural and electronic types in YBa2Cu3Oy near O half filling, ordered with respect to quench temperature, Tq, as a variable self-doping agent. On preparation through slow cooling to 260K pure orthorhombic alternate chain structure OI or O42 forms based on (4) and (2). (3) is generated with rising temperature according to equilibrium1 (4)+(2) =2(3), presumably in ordered infinitely adaptable ladder modifications. (3) is initially non-doping. At a critical %(3), near 600K, cell volume starts to rise strongly. At 673K and y=6.44, a cell volume expanded V+P+ variety optimizes, based now largely on (3). It, in tandem with apical $O^-$, is n-doping the planes and exhibits a Tc=100K level. Small amounts of a Tc=200K* level hint at the presence of ETS. Laser pulsing can also induce ETS, which is explained by (2), acting in tandem with $2O^-$, as a plane reducer and forming hole-pairs of its own. At yet higher quench temperatures tendencies are to tetragonal semiconductors based on (3) with V0<V+.

| Tq (K) | y=6.44 | Type | Doped states | Tc level (K) |
|--------|--------|------|--------------|--------------|
| 260 | y=6.50 | O24 V- | $(5)^{Ox}/(4)^{Red}$ | 50 |
| 600 | | O243 | $(5)^{Ox}/(4)^{Red},(3)^{Ox}O^-$ | 0 |
| 673 | | O32 V+P+ | $(5)^{Red}/(3)^{Ox}O^-$ | 100 |
| 673 minority | | | $(5)^{Red}/(2)^{Ox}2O^-$ | 200* |
| laser | y=6.50 | O24 C-P+ | $(5)^{Red}/(2)^{Ox}2O^-$ | 550* |